%% file: ms.tex
\documentclass[preprint,12pt]{aastex}

\shortauthors{Winn et al.~2008}
\shorttitle{Transit of HD~17156b}

\begin{document}

%
\def\ltsima{$\; \buildrel < \over \sim \;$}
\def\lsim{\lower.5ex\hbox{\ltsima}}
\def\gtsima{$\; \buildrel > \over \sim \;$}
\def\gsim{\lower.5ex\hbox{\gtsima}}
%

\bibliographystyle{apj}

\title{
  The Transit Light Curve Project.\\
  X.~A Christmas Transit of HD~17156b }

\author{
Joshua N.\ Winn\altaffilmark{1},
Matthew J.\ Holman\altaffilmark{2},
Gregory W.\ Henry\altaffilmark{3},
Guillermo Torres\altaffilmark{2},\\
Debra Fischer\altaffilmark{4},
John Asher Johnson\altaffilmark{5,6},
Geoffrey W.~Marcy\altaffilmark{7},
Avi Shporer\altaffilmark{8},
Tsevi Mazeh\altaffilmark{8}
}

\altaffiltext{1}{Department of Physics, and Kavli Institute for
  Astrophysics and Space Research, Massachusetts Institute of
  Technology, Cambridge, MA 02139, USA}

\altaffiltext{2}{Harvard-Smithsonian Center for Astrophysics, 60
  Garden Street, Cambridge, MA 02138, USA}

\altaffiltext{3}{Center of Excellence in Information Systems,
  Tennessee State University, 3500 John A.\ Merritt Blvd., Box 9501,
  Nashville, TN 37209, USA}

\altaffiltext{4}{Department of Physics and Astronomy, San Francisco
  State University, San Francisco, CA 94132}

\altaffiltext{5}{Institute for Astronomy, University of Hawaii,
  Honolulu, HI 96822}

\altaffiltext{6}{NSF Astronomy \& Astrophysics Postdoctoral Fellow}

\altaffiltext{7}{Department of Astronomy, University of California,
  Mail Code 3411, Berkeley, CA 94720}

\altaffiltext{8}{Wise Observatory, Tel Aviv University, Tel Aviv,
  Israel 69978}

\begin{abstract}

  Photometry is presented of the Dec.~25, 2007 transit of HD~17156b,
  which has the longest orbital period and highest orbital
  eccentricity of all the known transiting exoplanets. New
  measurements of the stellar radial velocity are also presented. All
  the data are combined and integrated with stellar-evolutionary
  modeling to derive refined system parameters. The planet's mass and
  radius are found to be $3.212_{-0.082}^{+0.069}$~$M_{\rm Jup}$ and
  $1.023_{-0.055}^{+0.070}$~$R_{\rm Jup}$. The corresponding stellar
  properties are $1.263_{-0.047}^{+0.035}$~$M_\sun$ and
  $1.446_{-0.067}^{+0.099}$~$R_\odot$. The planet is smaller by
  1$\sigma$ than a theoretical solar-composition gas giant with the
  same mass and equilibrium temperature, a possible indication of
  heavy-element enrichment. The midtransit time is measured to within
  1~min, and shows no deviation from a linear ephemeris (and therefore
  no evidence for orbital perturbations from other planets). We
  provide ephemerides for future transits and superior
  conjunctions. There is an 18\% chance that the orbital plane is
  oriented close enough to edge-on for secondary eclipses to occur at
  superior conjunction. Observations of secondary eclipses would
  reveal the thermal emission spectrum of a planet that experiences
  unusually large tidal heating and insolation variations.

\end{abstract}


\section{Introduction}

It is possible to estimate the mass and radius of a transiting planet
using a combination of photometry, Doppler data, and stellar modeling,
just as has long been done for eclipsing binary stars (Vogel
1890). Some aspects of the planetary orbit and atmosphere can also be
measured through high-precision transit photometry and spectroscopy
(see, e.g., Charbonneau et al.~2007, Seager 2008, Winn 2008). These
opportunities are more likely to occur for short-period planets,
because the probability for a randomly-oriented orbit to be viewed
close enough to edge-on for transits declines as $P^{-2/3}$. This
explains why all but one of the known transiting planets have periods
smaller than 10~days.

The exception is HD~17156b, for which $P=21.2$~days. This planet was
discovered in a Doppler survey by Fischer et al.~(2007). The
probability for transits to occur was larger than one might have
guessed based only on the period, because the planet is near
pericenter at the time of inferior conjunction (see, e.g., Burke~2008,
Barnes~2008, or Kane \& von Braun~2008), and indeed transits were
discovered by Barbieri et al.~(2007). The relatively long period,
along with the large orbital eccentricity of 0.67, presents an
interesting opportunity to study the planetary atmospheric response to
strongly time-variable heating by the parent star (Langton \& Laughlin
2007). However, the long period also presents a challenge to
observers. From a given site, there are only 2-3 good opportunities
each year to observe a complete transit. Irwin et al.~(2008), Narita
et al.~(2008), and Gillon et al.~(2008) observed the photometric
transits of Oct.~1, Nov.~12, and Dec.~3, 2007, respectively. They were
able to improve upon the precision of the system parameters given
originally by Barbieri et al.~(2007), but left further scope for
improvement through higher-precision photometry.

In this paper we present photometry of the transit of Dec.~25, 2007
based on observations with 4 different telescopes. We also present 10
new measurements of the Doppler shift of the host star, gathered
outside of transits. We analyze all of these data to refine the
estimates of the system parameters, using similar techniques to those
we have applied previously as part of the Transit Light Curve (TLC)
project (Holman et al.~2006, Winn et al.~2007) and that were recently
applied to a sample of 23 planets by Torres et al.~(2008). The
observations and data reduction are described in \S~2. The data
analysis is described in \S~3. In \S~4, the results of the data
analysis are used together with other observed stellar properties and
stellar evolutionary models to determine the properties of the star
and planet. All of the results are summarized in \S~5, and potential
future studies are discussed.

\section{Observations and Data Reduction}

\subsection{Out-of-transit radial velocities}

Fischer et al.~(2007) reported 33 measurements of the radial velocity
(RV) of HD~17156 over a time range from January 2006 to February
2007. Of these, 9 velocities were obtained with the Subaru~8m
telescope and HDS spectrograph, with a precision of approximately
5~m~s$^{-1}$. The other 24 velocities were based on observations with
the Keck~I 10m telescope and HIRES spectrograph (Vogt et al.~1994),
with a precision of 1-2~m~s$^{-1}$.

To these, we add 10 new Keck/HIRES velocities with 1-2~m~s$^{-1}$
precision that were obtained between August 2007 and March 2008. The
new data are based on observations with the same telescope,
instrument, and setup as the previous Keck observations. In particular
we employed an iodine gas absorption cell to calibrate the
instrumental profile and wavelength scale. To maintain a consistent
signal-to-noise ratio of about 200 per resolution element, we employed
the HIRES exposure meter, which uses a pickoff mirror to direct a
small fraction of the starlight to a photomultiplier tube and monitors
the exposure level as a function of time (Kibrick et al.~2006).

\begin{figure}[p]
\epsscale{1.0}
\plotone{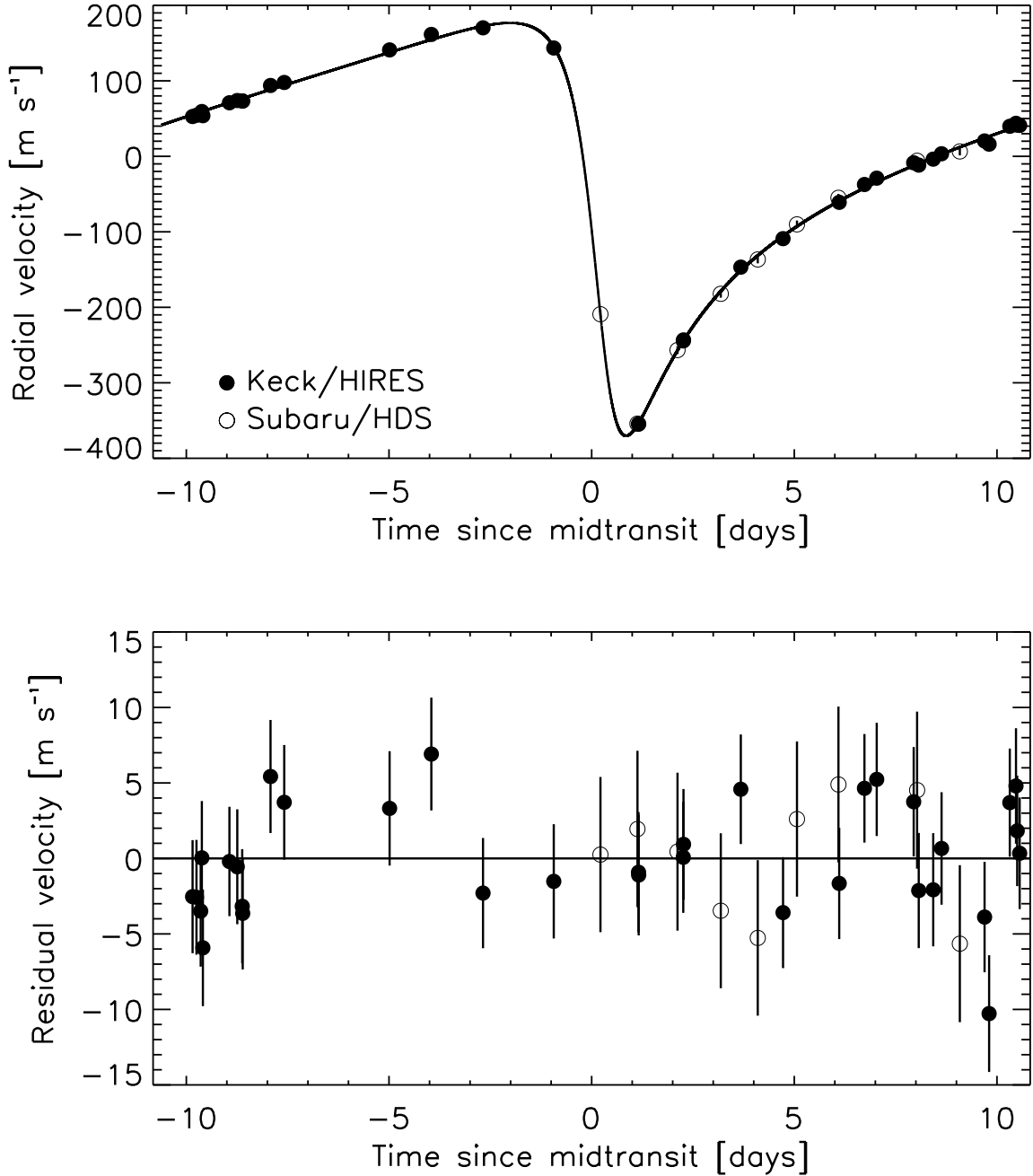}
\caption{ Radial velocity (RV) variation of HD~17156. {\it Top.}---The
measured RVs as a function of orbital phase, expressed
in days since midtransit. {\it Bottom.}---Differences between
the observed and calculated RVs, using the model described
in \S~3.3. The root-mean-squared (RMS) residual is 3.8~m~s$^{-1}$.
\label{fig:rv}}
\end{figure}

All 34 of the Keck/HIRES velocities were re-measured based on an
improved reduction of the raw CCD images and a refined version of the
algorithm of Butler et al.\ (1996), including the use of a new stellar
template. (The stellar template, an important ingredient in the
deconvolution algorithm of Butler et al.~1996, is an observation of
the target star obtained without the iodine cell and with a higher
signal-to-noise ratio and higher resolution than the rest of the
spectra.) The resulting velocities are given in
Table~\ref{tbl:rv}. For convenience, this table also includes the
Subaru/HDS velocities reported previously. Fig.~\ref{fig:rv} shows the
radial velocities as a function of orbital phase, using the ephemeris
derived in \S~3.

\begin{deluxetable}{ccr@{.}lc}

\tabletypesize{\scriptsize}
\tablecaption{Relative Radial Velocities (RV) of HD~17156\label{tbl:rv}}
\tablewidth{0pt}

\tablehead{
\colhead{Observatory Code\tablenotemark{a}} &
\colhead{Heliocentric Julian Date} &
\multicolumn{2}{c}{RV~[m~s$^{-1}$]} &
\colhead{Measurement Uncertainty~[m~s$^{-1}$]}
}

\startdata
\input table_rv.tex
\enddata

\tablenotetext{a}{
(1) Keck/HIRES.
(2) Subaru/HDS.
}

\tablecomments{The time stamps represent the Heliocentric Julian Date
  at the time of midexposure.}

\end{deluxetable}

\subsection{Transit photometry}

We observed the transit of UT~25~December~2007 (JD~2454459) using
telescopes at two different observatories in Arizona: the Fred
L.~Whipple Observatory (FLWO) and Fairborn Observatory. The Moon was
full. The weather was generally clear over both observatories,
although there were some light clouds and transparency variations.

At FLWO, we used the 1.2m telescope and Keplercam, a $4096^2$ CCD with
a $23\arcmin \times 23\arcmin$ field of view (Szentgyorgyi et
al.~2005). The images were binned $2\times 2$, giving a scale of
$0\farcs 68$ per binned pixel. We obtained 10~s exposures though a
$z$-band filter for 6.5~hr bracketing the predicted midtransit
time. The telescope was defocused to avoid significant nonlinearity
and saturation, giving a typical full width at half-maximum (FWHM) of
stellar images of 4$\arcsec$ (6~pixels). The time between exposures
was 11~s due to readout and reset operations. Autoguiding failures led
to pointing drifts of 10~pixels in right ascension and 40~pixels in
declination over the course of the night. During the observations the
target rose from airmass 1.34 to 1.30 (reaching the meridian at
HJD~2454459.67) and then set to airmass 1.64.

We used standard IRAF procedures for overscan correction, trimming,
bias subtraction, flat-field division, and aperture photometry of
HD~17156 and 32 other stars in the field of view (all necessarily
fainter than HD~17156). We created a reference signal by combining the
normalized light curves of the different comparison stars, and divided
the flux history of HD~17156 by this reference signal. We experimented
with different choices for the aperture size, combinations of the
comparison stars, and weighting schemes for combining the normalized
light curves, aiming to minimize the standard deviation of the
out-of-transit portion of the light curve. Best results were obtained
when the aperture diameter was 15~pixels and the comparison light
curves were weighted by the square root of the mean flux.

At Fairborn Observatory, we used 3 independent telescopes: the T8,
T10, and T11~0.8m automated photometric telescopes (APTs). All of the
APTs are equipped with two temperature-stabilized EMI~9124QB
photomultiplier tubes for measuring photon count rates simultaneously
through Str\"omgren $b$ and $y$ filters. Each telescope nodded back
and forth between HD~17156 ($V=8.17$, $B-V=0.64$) and the comparison
star HD~15784 ($V=6.64$, $B-V=0.64$), which was found to be constant
in brightness by Fischer et al.~(2007). The T8, T10, and T11 APTs
obtained 170, 190, and 200 observations of both stars, respectively,
over a period 7.2 hours. From these measurements, we computed a total
of 560 target-minus-comparison differential magnitudes in each $b$ and
$y$ photometric band. The time series was trimmed to 523 points to
eliminate bad data taken at the highest airmass. To increase the
signal-to-noise ratio in the resulting light curves, we averaged the
$b$ and $y$ passbands together, resulting in a synthetic $(b+y)/2$
passband.

\begin{figure}[p]
\epsscale{1.0}
\plotone{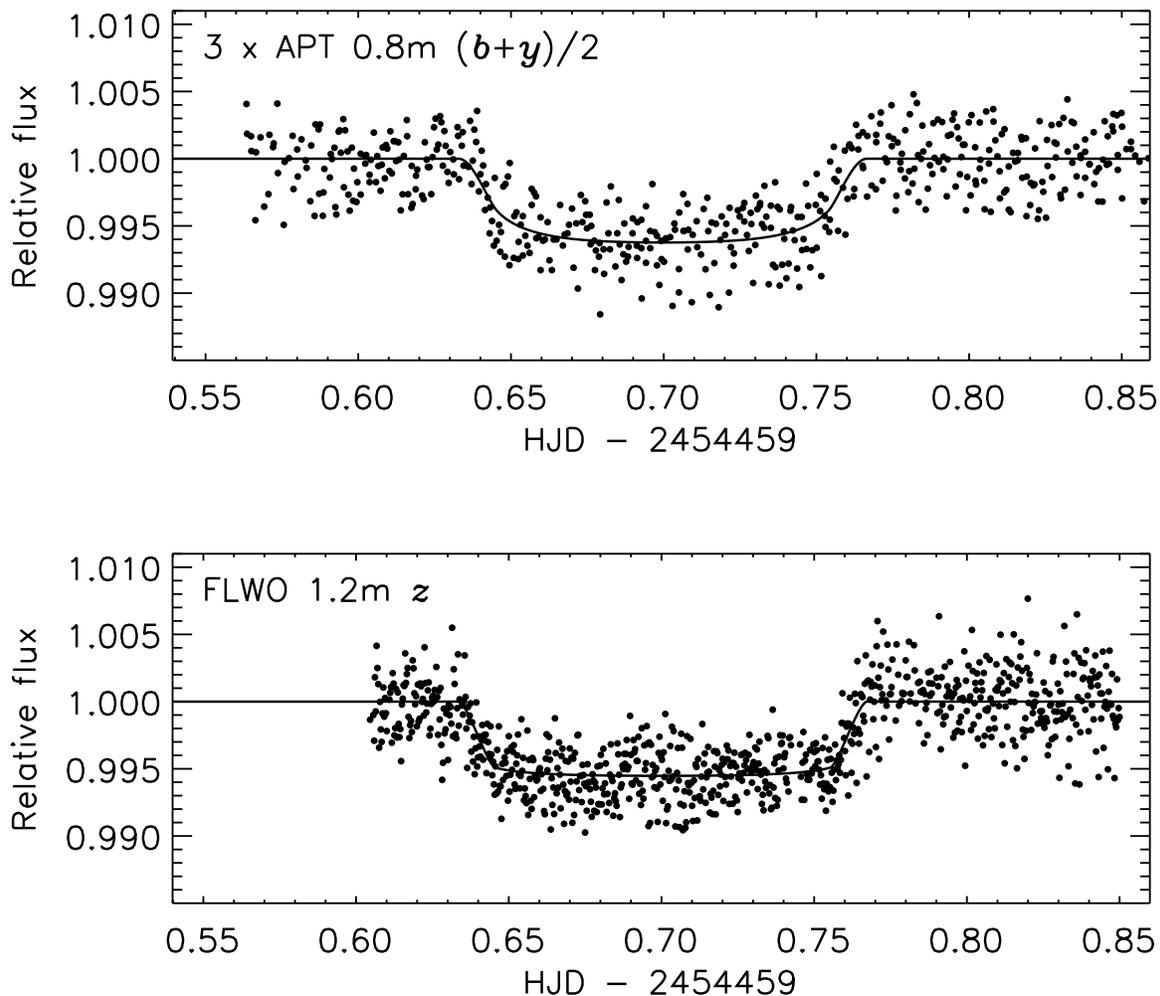}
\caption{ Relative photometry of HD~17156 on UT~25~Dec~2007, along
  with the best-fitting model.
{\it Top}.---Data from the T8, T10, and
  T11~0.8m automated photometric telescopes at Fairborn Observatory,
  in the ``$(b+y)/2$'' band (the Str\"omgren $b$ and $y$ data were
  averaged). The median time interval between data points is 43~s. The
  standard deviation of the out-of-transit (OOT) data and of the residuals are 0.00214
  and 0.00213, respectively.
{\it Bottom}.---Data from the Fred
  L.~Whipple 1.2m telescope and Keplercam, in the $z$ band. The median
  time interval between data points is 24~s. The standard deviation of
  the OOT data and of the residuals are 0.00233 and 0.00206, respectively.
\label{fig:phot}}
\end{figure}

The FLWO and APT light curves were corrected for differing airmass
extinction between the target star and comparison stars, by fitting
the observed magnitudes to a linear function of airmass (see
Eqn.~\ref{eq:extinction}). The parameters of the linear function were
determined simultaneously with the other model parameters, as
described in \S~3. The extinction-corrected data are given in
Table~\ref{tbl:phot}, and plotted in Fig.~\ref{fig:phot}, along with
the best-fitting model. For the APT data, the standard deviation of
the out-of-transit (OOT) data and of the residuals are 0.00214 and
0.00213, respectively.  The corresponding figures for the FLWO data
are 0.00233 and 0.00206, respectively. For both light curves, the
expected photometric precision due only to Poisson noise and
scintillation noise (using the empirical formulas of Reiger~1963,
Young~1967 and Dravins~1998) is about 50\% smaller than the OOT
standard deviation, with the dominant contribution arising from
scintillation noise. It is possible that the scintillation noise was
higher than the empirical and highly approximate formula predicts. It
is also likely that there are additional noise sources such as
transparency variations beyond the simple dependence on airmass.

\begin{deluxetable}{ccc}

\tabletypesize{\scriptsize}
\tablecaption{Relative Photometry of HD~17156\label{tbl:phot}}
\tablewidth{0pt}

\tablehead{
\colhead{Observatory Code\tablenotemark{a}} &
\colhead{Heliocentric Julian Date} &
\colhead{Relative flux}
}

\startdata
  $  1$  &  $  2454459.56340$  &  $  1.0041$ \\
  $  1$  &  $  2454459.56360$  &  $  1.0019$ \\
  $  1$  &  $  2454459.56490$  &  $  1.0017$ \\
  $  1$  &  $  2454459.56500$  &  $  1.0006$ \\
  $  1$  &  $  2454459.56630$  &  $  0.9954$ \\
  $  1$  &  $  2454459.56650$  &  $  1.0005$ \\
  $  1$  &  $  2454459.56800$  &  $  1.0016$ \\
  $  1$  &  $  2454459.56920$  &  $  0.9964$ \\
  $  1$  &  $  2454459.57060$  &  $  0.9974$ \\
  $  1$  &  $  2454459.57090$  &  $  1.0018$ \\
  $  1$  &  $  2454459.57210$  &  $  1.0009$ \\
  $  1$  &  $  2454459.57230$  &  $  1.0011$
\enddata

\tablenotetext{a}{
(1) T8, T10, and T11 APT 0.8m telescopes, Fairborn Observatory, Arizona, USA.
(2) Fred L.~Whipple Observatory 1.2m telescope, Arizona, USA.
}

\tablecomments{The time stamps represent the Heliocentric Julian Date
  at the time of mid-exposure. We intend for this Table to appear in
  entirety in the electronic version of the journal. An excerpt is
  shown here to illustrate its format. The data are also available
  from the authors upon request.}

\end{deluxetable}

\section{Analysis of radial-velocity and photometric data}

At the heart of our analysis was a simultaneous fit of a parametric
model to the new light curves and the available radial-velocity
data. In \S~3.1 we describe the model and fitting procedures. In
\S~3.2 we provide more detail about the photometric noise and how it
was modeled.  In \S~3.3 we discuss the determination of midtransit
times and a new transit ephemeris. In \S~4 we explain how the data
analysis was integrated with stellar evolutionary models to determine
the final system parameters.

\subsection{Joint radial velocity and photometric analysis}

The model is based on a two-body Keplerian orbit, with the loss of
light during transits given by the formulae of Mandel \&
Agol~(2002). The orbit is parameterized by the semi-amplitude ($K$),
period ($P$), midtransit time ($T_c$)\footnote{Since the orbit of
  HD~17156b is highly eccentric, the definition of $T_c$ requires some
  care. Here we define it as the time when the projected planet-star
  separation is smallest, which is also the time of minimum light for
  a limb-darkened star. This is to be distinguished from other
  possible definitions, such as the halfway point between first and
  last contact, or the moment when the true anomaly $f$ of the
  planetary orbit is equal to $\pi/2-\omega$.}, inclination ($i$),
eccentricity ($e$), and argument of pericenter ($\omega$). The
coordinate system is chosen such that $i<90^\circ$ and the longitude
of nodes is zero.

Additional parameters relevant to the photometric data are the
planet-to-star radius ratio ($R_p/R_\star$), the semimajor axis in
units of the stellar radius ($a/R_\star$), and the limb-darkening
coefficients ($u_1$ and $u_2$) for each light curve. We assumed a
quadratic limb-darkening law,
\begin{equation}
\frac{I_\mu}{I_1} = 1 - u_1(1-\mu) - u_2(1-\mu)^2,
\end{equation}
where $\mu$ is the cosine of the angle between the line of sight and
the normal to the stellar surface, and $I_\mu$ is the specific
intensity. In many studies of transiting planets, the limb-darkening
coefficients are often held fixed at values deemed appropriate for the
host star, based on stellar-atmosphere models. However, when the data
are sufficiently precise it is preferable to fit for the coefficients,
as recently emphasized by Southworth~(2008). This is because the model
atmospheres might be wrong, and at least are uncertain to some
degree. Neglecting this uncertainty leads to underestimated errors in
all parameters that are covariant with the limb-darkening
coefficients, namely $R_p/R_\star$, $R_\star/a$, and $i$. For this
study we allowed $u_1$ and $u_2$ to vary freely subject only to the
conditions $u_1+u_2<1$ (nonnegative intensity at the limb),
$u_1+u_2>0$ (fainter at the limb than the center), and $u_1>0$
(maximum intensity at the center). It proved advantageous to perform
the fit using the linear combinations
\begin{equation}
v_1 = u_1 + \frac{7}{5}u_2,~~v_2 = -\frac{7}{5}u_1 + u_2,
\end{equation}
because $v_1$ and $v_2$ have nearly uncorrelated errors (for further
discussion, see P\'al~2008).

The model also has 6 nuisance parameters. For each of the two RV data
sets there is an additive constant velocity ($\gamma_K$ for Keck/HIRES
and $\gamma_S$ for Subaru/HDS). For each of the two light curves there
are the parameters $m_0$ and $k$ of the differential extinction
correction,
\begin{equation}
\label{eq:extinction}
m_{\rm cor} = m_{\rm obs} + m_0 + kz,
\end{equation}
where $m_{\rm obs}$ is the observed magnitude, $m_{\rm cor}$ is the
corrected magnitude, and $z$ is the airmass.

All together there were 18 parameters: $K$, $P$, $T_c$, $i$, $e$,
$\omega$, $R_p/R_\star$, $a/R_\star$, an additive velocity for each of
two RV data sets, two limb-darkening coefficients $u_1$ and $u_2$ for
each of two bandpasses, and two parameters $m_0$ and $k$ for each of
two bandpasses to describe the correction for differential airmass
extinction. The fitting statistic was
\begin{equation}
\label{eq:chi2}
\chi^2 = \chi^2_F + \chi^2_V + \chi^2_T,
\end{equation}
with the terms defined as follows. The first term is based on the fit
to the photometric data:
\begin{equation}
\chi^2_F = \sum_{i=1}^{1372} \left[ \frac{f_i({\rm obs}) - f_i({\rm calc})}{\sigma_{f,i}} \right]^2,
\label{eq:chi2_flux}
\end{equation}
where $f_i$(obs) is the $i$th measured flux (photometric data point),
$f_i$(calc) is the calculated flux given a particular choice of model
parameters, and $\sigma_{f,i}$ is the uncertainty in the $i$th
measured flux. The choice of $\sigma_{f,i}$ is nontrivial and is
discussed in \S~3.2. Here we simply state our choice to take
$\sigma_{f,i}$ to be a constant for each light curve given by
$\sigma_{f,i} = \beta \sigma_{\rm res}$, where $\sigma_{\rm res}$ is
the standard deviation of the flux residuals, and $\beta \geq 1$ is a
factor intended to account for time-correlated errors. For the APT
light curve, $\sigma_{\rm res} = 0.00213$ and $\beta=1.03$, giving an
effective uncertainty per point of 0.0022.  For the FLWO light curve,
$\sigma_{\rm res} = 0.00207$ and $\beta=1.28$, giving an effective
uncertainty per point of 0.0026.

The second term in Eqn.~(\ref{eq:chi2}) is based on the fit to the RV
data:
\begin{equation}
\chi^2_V = \sum_{i=1}^{43} \left[ \frac{V_i({\rm obs}) - V_i({\rm calc})}{\sigma_{V,i}} \right]^2,
\end{equation}
where $V_i$(obs) and $V_i$(calc) are the $i$th observed and calculated
RVs, and $\sigma_{V,i}$ is the corresponding uncertainty.  For
$\sigma_{V,i}$, we used the quadrature sum of the measurement
uncertainty given in Table~\ref{tbl:rv} and a constant $\sigma_{V,0} =
3.4$~m~s$^{-1}$ intended to account for stellar ``jitter,'' excess
noise that is usually attributed to motions of the stellar
photosphere. This choice resulted in a reduced $\chi^2_V$ of unity
when fitting the RV data only, and it is consistent with observations
of other stars of similar type (Wright 2005).

The third term in Eqn.~(\ref{eq:chi2}) is an {\it a priori}\,
constraint enforcing the transit ephemeris that is derived in \S~3.3.
Specifically,
\begin{equation}
\label{eq:chi2-t}
\chi^2_T = \left[ \frac{P~{\rm (days)} - 21.21688}{0.00044} \right]^2 +
\left[ \frac{T_c~{\rm (HJD)} - 2,454,459.69987}{0.00045} \right]^2.
\end{equation}
The central values of $P$ and $T_c$, and their uncertainties, are
derived in \S~3.3 based on observations of 5 different transits
spanning 106~days. They are more precise than could be derived
internally from only the RV data, the APT data, and the FLWO data that
are fitted here; hence during this step we treated $P$ and $T_c$ as
externally measured quantities. Since the transit ephemeris is based
in part on the midtransit times of the APT and FLWO light curves, we
used an iterative procedure: first, a previously published ephemeris
was used in Eqn.~(\ref{eq:chi2-t}); second, the midtransit times based
on the APT and FLWO light curves were determined as in \S~3.3; and
third, the ephemeris was re-derived and used in the next iteration of
the fit to the RV and photometric data. Further iterations were
performed but made no appreciable difference in the results.

We used a Markov Chain Monte Carlo (MCMC) algorithm to estimate the
best-fitting model parameters and their uncertainties (see, e.g.,
Appendix A of Tegmark et al.~2004). This algorithm creates a sequence
of points (a ``chain'') in parameter space by iterating a jump
function, which in our case was the addition of a Gaussian random
deviate to a randomly-selected single parameter. After this operation,
if the new point has a lower $\chi^2$ than the previous point, the
``jump'' is executed: the new point is added to the chain. If not,
then the jump is executed with a probability proportional to
$\exp(-\Delta\chi^2/2)$. If the jump is not executed, the current
point is repeated in the chain. The sizes of the random deviates are
set to values for which $\sim$40\% of jumps are executed. After
creating multiple chains to check for mutual convergence, and trimming
off the initial segments to eliminate artifacts of the initial
condition, the density of the chain's points in parameter space is
taken to be the joint {\it a posteriori}\, probability distribution of
the parameter values. Probability distributions for individual
parameters are created by marginalizing over all other parameters. For
each parameter we report the mode of the distribution and the 68.3\%
confidence limits, defined by the 15.85\% percentile and the 84.15\%
percentile in the cumulative distribution.

The results are given in Table~\ref{tbl:params}, with the designation
A in the last column. The entries designated B are those that are
drawn from other works and are repeated here for convenience.  The
entries designated C are based on a synthesis of our modeling results
and theoretical models of stellar evolution, as discussed in
\S~4. The results for the limb-darkening parameters are given
separately, in Table~\ref{tbl:ld}.

\begin{deluxetable}{lccc}
\tabletypesize{\scriptsize}
\tablecaption{System Parameters of HD~17156\label{tbl:params}}
\tablewidth{0pt}

\tablehead{
\colhead{Parameter} & \colhead{Value} & \colhead{68.3\% Conf.~Limits} & \colhead{Comment}
}

\startdata
{\it Transit and orbital parameters:} & & & \\
Orbital period, $P$~[d]                               & $21.21688$       &  $\pm 0.00044$     & A  \\
Midtransit time~[HJD]                                 & $2454459.69987$  &  $\pm 0.00045$    & A  \\
Planet-to-star radius ratio, $R_p/R_\star$             & $0.0727$         &  $\pm 0.0016$     & A  \\
Orbital inclination, $i$~[deg]                        & $86.2$           &  $-0.8$, $+2.1$  & A  \\
Scaled semimajor axis, $a/R_\star$                     & $22.8$           &  $-1.7$, $+2.2$  & A  \\
Transit impact parameter, $b_{\rm I}$                   & $0.55$           &  $-0.29$, $+0.03$ & A  \\
Transit duration~[hr]                                 & $3.177$          &  $-0.041$, $+0.071$ & A  \\
Transit ingress or egress duration~[hr]               & $0.25$          &  $-0.026$, $+0.078$ & A  \\
Velocity semiamplitude, $K$~[m~s$^{-1}$]              & $272.7$          &  $\pm 2.1$       & A  \\
Orbital eccentricity, $e$                             & $0.6753$         &  $\pm 0.0036$    & A  \\
Argument of pericenter, $\omega$~[deg]                & $121.64$         &  $\pm 0.48$      & A  \\
Planet-to-star mass ratio, $M_p/M_\star$               & $0.00244$        &  $\pm 0.000029$  & C \\
Semimajor axis~[AU]                                   & $0.1623$         &  $-0.0020$, $+0.0015$   & C \\
& & & \\
{\it Stellar parameters:} & & & \\
Mass, $M_\star$~[M$_{\odot}$]                           &  $1.263$         &  $-0.047$, $+0.035$  & C  \\
Radius, $R_\star$~[R$_{\odot}$]                         &  $1.446$         &  $-0.067$, $+0.099$  & C  \\
Surface gravity, $\log g_\star$~[cgs]                 &  $4.219$         &  $-0.055$, $+0.033$   & C  \\
Mean density, $\rho_\star$~[g~cm$^{-3}$]               &  $0.589$         &  $-0.103$, $+0.066$   & A  \\
Effective temperature, $T_{\rm eff}$~[K]               &  $6079$          &  $\pm 80$      & B  \\
Metallicity, [Fe/H]                                  &  $+0.24$         &  $\pm 0.05$    & B  \\
Projected rotation rate, $v\sin i_\star$~[km~s$^{-1}$]  &  $2.6$          &  $\pm 0.5$     & B  \\
Luminosity [L$_\odot$]                                &  $2.55$          &  $-0.32$, $+0.24$   & C  \\
Absolute $V$ magnitude                                &  $3.78$          &  $-0.27$,$+0.23$  & C \\
Age [Gyr]                                             &  $3.06$          &  $-0.76$, $+0.64$  & C  \\
& & & \\
{\it Planetary parameters:} & & & \\
$M_p$~[M$_{\rm Jup}$]                                 &  $3.212$          &  $-0.082$,$+0.069$  & C \\
$R_p$~[R$_{\rm Jup}$]                                 &  $1.023$         &  $-0.055$,$+0.070$ & C \\
Surface gravity, $g_p$~[m~s$^{-2}$]                   &  $76.1$         &  $-9.0$,$+7.3$ & A  \\
Mean density, $\rho_p$~[g~cm$^{-3}$]                  &  $3.72$          &  $\pm 0.67$ & C  \\
\enddata

\tablecomments{ (A) Based on the joint analysis of photometric and RV
  data (see \S~3.1-3.3). (B) From Fischer et al.~(2007), with enlarged
  error bars for $T_{\rm eff}$ and [Fe/H]. (C) Functions of group A
  and B parameters, supplemented as needed by an isochrone analysis
  (see \S~4), the Tycho-2 apparent magnitudes (H{\o}g et al.~2000),
  and {\it Hipparcos}\, parallax of $\pi=13.34\pm 0.72$~mas (van
  Leeuwen 2007). }

\end{deluxetable}

\begin{deluxetable}{lccc}
\tabletypesize{\scriptsize}
\tablecaption{Fitted Limb-Darkening Coefficients for HD~17156\label{tbl:ld}}
\tablewidth{0pt}

\tablehead{
\colhead{Bandpass} & \colhead{Linear Coefficient ($u_1$)} & \colhead{Quadratic Coefficient ($u_2$)}
}

\startdata
$z$       & $0.13_{-0.06}^{+0.30}$ & $0.00_{-0.13}^{+0.25}$ \\
$(b+y)/2$ & $0.21_{-0.10}^{+0.42}$ & $0.64_{-0.51}^{+0.13}$
\enddata

\end{deluxetable}

\subsection{Photometric noise analysis}

Deriving reliable uncertainties in the system parameters requires a
realistic treatment of the photometric noise. In this work, as in
others, we have based our analysis on a $\chi^2$ statistic that
implicitly treats the noise as independent and identically-distributed
Gaussian random variables added to each datum. This is because simple
and fast statistical methods are applicable in that case, and because
there is no clear alternative. Unfortunately, precise photometry is
often plagued with time-correlated (``red'') noise, which is probably
responsible for the frequent disagreements between reported values of
photometric parameters in the literature. It is advisable to attempt
to justify the common assumption of independent Gaussian noise, or at
least to gain some understanding of the limitations of that
assumption.\footnote{In this vein we quote an email from G.~Kovacs:
  ``Identifying the `red noise component' (whatever it means) in a
  single realization of a rather limited time series (such as
  photometric followup data) is a mission impossible.''}.

\begin{figure}[p]
\epsscale{0.8}
\plotone{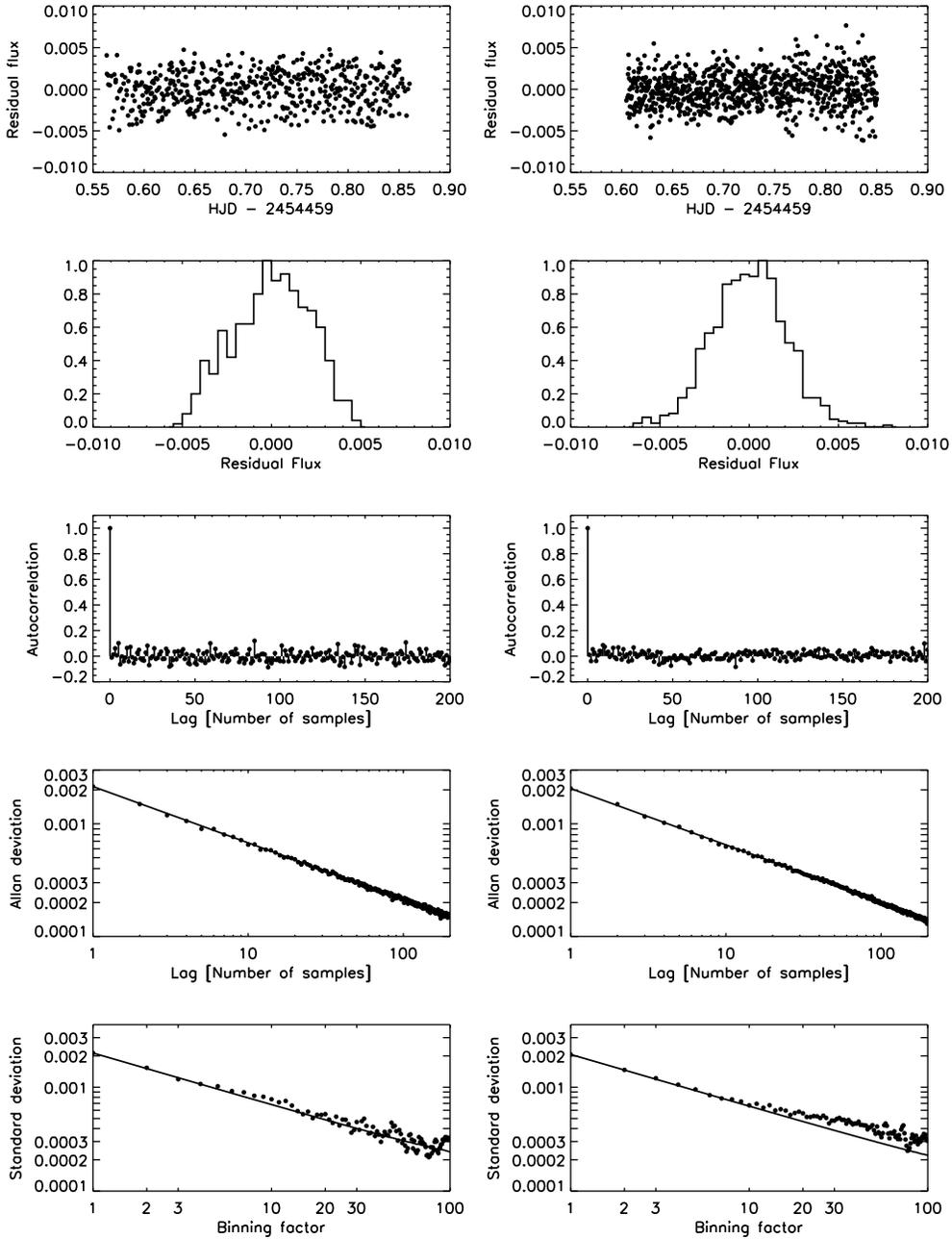}
\caption{ Photometric noise analysis. The left column
applies to the APT $(b+y)/2$ data, and the right column
applies to the FLWO~1.2m $z$ data.
{\it First row.}---Residuals from the best-fitting
model (observed$-$calculated).
{\it Second row.}---Histogram of the residuals.
{\it Third row.}---Autocorrelation of the residuals
as a function of the lag.
{\it Fourth row.}---Allan deviation of the residuals,
as a function of the lag. The straight line shows
the $l^{-1/2}$ dependence that is expected of
white Gaussian noise.
{\it Fifth row.}---Standard deviation of time-binned
residuals, as a function of the number of data points
per bin. The straight line shows the expectation
for white Gaussian noise (see Eqn.~\ref{eq:beta}).
\label{fig:noise}}
\end{figure}

To investigate the noise in our light curves, we performed some tests
on the residuals. The top two panels in Fig.~\ref{fig:noise} show the
residuals for the two independent light curves. The second row shows
histograms of the residuals, which are approximately Gaussian. The
third row shows the autocorrelation of the residuals as a function of
the lag,
\begin{equation}
A(l) = \frac
{\sum_{i=0}^{N-l-1} (r_i-\bar{r})(r_{i+l}-\bar{r})}
{\sum_{i=0}^{N-1} (r_i-\bar{r})^2},
\end{equation}
where $N$ is the number of data points, $r_i$ is the $i$th residual
flux (measured flux minus calculated flux) and $\bar{r}$ is the mean
residual flux.  The autocorrelations are $\lesssim$0.1. The fourth row
shows the Allan deviation $\sigma_A(l)$ of the residuals, where
$\sigma_A$ is defined by (Allan 1964; Thompson, Moran, \& Swensen
2001)
\begin{equation}
\sigma_A^2(l) = \frac{1}{2(N+1-l)} \sum_{n=0}^{N-2l} \left[
\frac{1}{l} \sum_{m=0}^{l} \left( r_{n+m}-r_{n+l+m} \right)
\right]^2,
\end{equation}
where $N$ is the number of data points, $r_i$ denotes the $i$th
residual, and $l$ is the lag. The Allan deviation is commonly used in
the time metrology literature to assess 1/$f$ noise. The solid lines
show the dependence $\sigma_A\propto l^{-1/2}$ that is expected of
white Gaussian noise. The data fall close to these lines. The bottom
row of panels shows the standard deviation of the time-binned
residuals, as a function of the number of data points per
bin. Specifically, we averaged the residuals into $m$ bins of $n$
points and calculated the standard deviation $\sigma_n$ of the binned
residuals. In the absence of red noise, one would expect
\begin{equation}
\label{eq:beta}
\sigma_n = \frac{\sigma_1}{n^{1/2}} \left( \frac{m}{m-1} \right)^{1/2},
\end{equation}
but in reality $\sigma_n$ is larger than this expression by a factor
$\beta$. In general $\beta$ depends on $m$ but the dependence is weak.
For bin sizes ranging from 10-30~min (the most important timescale of
the transit light curve) the mean value of $\beta$ is $1.03$ for the
APT light curve and $1.28$ for the FLWO light curve. Apparently the
APT residuals average down as one would expect of white noise, while
the FLWO residuals are correlated to some degree.

As stated above, our $\chi^2$-based method is strictly appropriate
only for uncorrelated errors. One might imagine modifying the
definition of $\chi^2_F$ to employ the full covariance matrix of the
errors rather than assuming independent errors. However, given that
there is little structure in the autocorrelation functions to provide
guidance on how to model the covariance matrix, and that the
correlations seem small, we account for correlated noise in a simple
and approximate fashion: we assign a photometric error bar of
$\sigma_{f,i} = \beta \sigma_{\rm res}$ to each data point, where
$\sigma_{\rm res}$ is the standard deviation of the residuals. The
choice $\beta=1$ is essentially equivalent the common procedure of
assuming equal errors for all data points and scaling the error bars
such that $\chi^2/N_{\rm dof} = 1$. We used $\beta=1.03$ for the APT
data and $\beta=1.28$ for the FLWO data.

\subsection{Midtransit times}
\label{sec:ttv}

Measurements of midtransit times are important for refining the
transit ephemeris and thereby enabling accurate predictions for future
observations, and also for searching for satellites and additional
planets via the method of Holman \& Murray~(2005) and Agol et
al.~(2005). We determined the midtransit time from each light curve
using 3 different methods to check for consistency: the MCMC algorithm
described in \S~3.1, and two different bootstrap analyses described
below. All of these methods rely on the photometric transit model and
the computation of $\chi^2_F$. Besides $T_c$, the only other variable
parameters were the slope ($k$) and offset ($m_0$) of the differential
extinction correction. These were the only other parameters covariant
with the midtransit time. All other parameters needed to specify the
transit model were held fixed at the best-fitting values.

The first bootstrap method was ``random draws with replacement''
(Press et al.\ 1992, p.~689). It involves minimizing $\chi^2_F$ as a
function of the parameters for $10^4$ synthetic data sets, each of
which has the same number of data points as the real data. Each entry
in a synthetic data set is a datum (a time stamp and relative flux)
drawn randomly from the real data set, with repetitions allowed. Thus,
a substantial fraction of the entries in each synthetic data set are
duplicated at least once and receive greater weight in $\chi^2_F$
sum. The idea is to estimate the noise properties of the data using
the observed data values themselves, rather than choosing models for
the underlying physical process and for the noise. However, an
underlying assumption is that the errors are uncorrelated and
identically distributed. The distribution of results for $T_c$ is
taken to be the probability density for the midtransit time.

The second bootstrap method was ``residual permutation.'' It is
similar to the method just described but the synthetic data sets are
created differently. The residuals of the best-fitting model are added
back to the model light curve after performing a cyclic permutation of
their time indices. With $N$ data points, one may create $N-1$
synthetic data sets in this manner. Another $N-1$ may be created by
inverting the time order of the residuals and then performing cyclic
permutations. The idea is again to use the data themselves to estimate
the noise properties, but this time without assuming that the errors
are uncorrelated. The correlations between residuals at different
times are preserved. The distribution of results for $T_c$ is taken to
be the probability density for the midtransit time. A disadvantage of
this method is that only $2(N-1)$ realizations can be generated
without further assumptions, and therefore the distribution is
relatively noisy.

The results from all 3 methods are given in
Table~\ref{tbl:compare-methods}. As before, the quoted values are the
modes of the probability distributions and the quoted error bars range
from the 15.85\% percentile to the 84.15\% percentile. The
distributions are symmetric in all cases, and are nearly Gaussian for
the MCMC and random-draws methods. The residual-permutation method
produced distributions with broader wings than a Gaussian
function. For each light curve, the results from all 3 methods are in
agreement within 0.15$\sigma$, where $\sigma$ is the error in the MCMC
method. The bootstrap method gave the smallest error bars, as one
might expect, given that the bootstrap method ignores correlated
noise. (The bootstrap-derived error bar is smaller than the
MCMC-derived error bar by approximately the ``red noise'' factor
$\beta$.) In what follows, we adopt the MCMC results for concreteness
and for consistency with our previous analyses. The error bars on the
midtransit times derived from the FLWO and APT light curves are
1.1~min and 0.9~min, and the difference between the results is
1.0~min. This level of agreement is a consistency check on the
accuracy of our error bars.

\begin{deluxetable}{lcc}
\tabletypesize{\scriptsize}
\tablecaption{Comparison of Methods for Measuring Midtransit Times\label{tbl:compare-methods}}
\tablewidth{0pt}

\tablehead{
\colhead{Analysis method} & \multicolumn{2}{c}{Midtransit time~[HJD]} \\
\colhead{} & \colhead{FLWO} & \colhead{APT}
}

\startdata
MCMC with $\beta>1$  &  $ 2454459.70044 \pm 0.00075 $  &  $2454459.69972 \pm 0.00065$ \\
Random draws with replacement  &  $ 2454459.70034 \pm 0.00054 $  &  $2454459.69974 \pm 0.00063$ \\
Residual permutation &  $ 2454459.70048 \pm 0.00090 $  &  $2454459.69981 \pm 0.00095$
\enddata

\end{deluxetable}

\begin{figure}[ht]
\epsscale{1.0}
\plotone{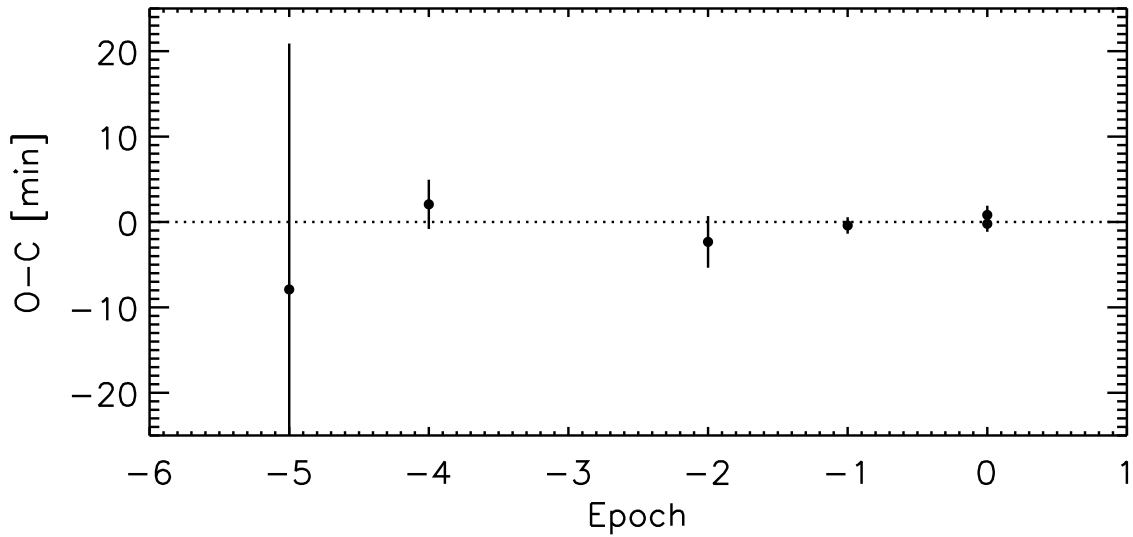}
\caption{Transit timing residuals for HD~17156b. The calculated times,
  using the ephemeris given in Eqns.~(\ref{eq:tc},\ref{eq:p}), have
  been subtracted from the observed times.
  \label{fig:ttv}}
\end{figure}

We fitted a linear ephemeris, $T_c[E] = T_c[0] + EP$, to all of the
midtransit times at our disposal: namely, the APT and FLWO midtransit
times presented in this paper, and the 4 different midtransit times
reported by Barbieri et al.~(2007), Gillon et al.~(2008), Irwin et
al.~(2008), and Narita et al.~(2008).  For convenience, all of the
midtransit times are given in Table~\ref{tbl:tc}.  The results were
\begin{eqnarray}
T_c[0] & = & 2,454,459.69987 \pm 0.00045~{\rm [HJD]}, \label{eq:tc} \\
P & = & 21.21688 \pm 0.00044~{\rm days} \label{eq:p}.
\end{eqnarray}
The fit gives $\chi^2 = 1.99$ with 4 degrees of freedom, suggesting
that a constant period is consistent with the available data. A plot
of the timing residuals (observed$-$calculated midtransit times) is
shown in Fig.~\ref{fig:ttv}. There are no obvious anomalies at the
level of a few minutes.

\begin{deluxetable}{ccc}
\tabletypesize{\scriptsize}
\tablecaption{Midtransit Times of HD~17156b\label{tbl:tc}}
\tablewidth{0pt}

\tablehead{
\colhead{Heliocentric Julian Date} & \colhead{1$\sigma$ Uncertainty} & \colhead{Reference}
}

\startdata
$2454353.61000$  &  $0.02000$ &  Barbieri et al.~(2007)  \\
$2454438.48271$  &  $0.00067$ &  Gillon et al.~(2008)  \\
$2454374.83380$  &  $0.00200$ &  Irwin et al.~(2008)  \\
$2454417.26450$  &  $0.00210$ &  Narita et al.~(2008)  \\
$2454459.70044$  &  $0.00075$ &  This work (FLWO) \\
$2454459.69972$  &  $0.00065$ &  This work (APT)
\enddata

\end{deluxetable}

\section{Theoretical isochrone fitting}

The RV and photometric data do not uniquely determine the masses and
radii of the planet and the star. Some external information about the
star or the planet must be introduced to break the fitting
degeneracies $M_p \propto M_\star^{2/3}$ and $R_p \propto R_\star
\propto M_\star^{1/3}$ (see, e.g., Winn 2008). We broke these
degeneracies by requiring consistency between the observed properties
of the star, the stellar mean density $\rho_\star$ that can be derived
from the photometric parameter $a/R_\star$ (Seager \& Mallen-Ornelas
2003, Sozzetti et al.~2007), and theoretical models of stellar
evolution. The inputs were $T_{\rm eff}=6079\pm 80$~K and
[Fe/H]~$=+0.24\pm 0.05$ (from Fischer et al.~2007 but with enlarged
error bars, as per Torres et al.~2008), the absolute
magnitude\footnote{The quoted $M_V$ is based on the transformation of
  Tycho-2 apparent magnitudes (H{\o}g et al.~2000) to the Johnson $V$
  band, giving $V=8.172\pm 0.012$, and the Hipparcos parallax of
  $\pi=13.34\pm 0.72$~mas (van Leeuwen~2007). We have assumed
  zero reddening.} $M_V=3.80\pm 0.12$,
the stellar mean density $\rho_\star =
0.589_{-0.103}^{+0.066}$~g~cm$^{-3}$ derived from the results for the
$a/R_\star$ parameter, and the Yonsei-Yale (Y$^2$) stellar evolution
models by Yi et al.~(2001) and Demarque et al.~(2004). We computed
isochrones for the allowed range of metallicities, and for stellar
ages ranging from 0.1 to 14 Gyr. For each stellar property (mass,
radius, and age), we took a weighted average of the points on each
isochrone, in which the weights were proportional to
$\exp(-\Delta\chi^2_\star/2)$ with
\begin{equation}
\chi^2_\star =
\left[ \frac{\Delta{\rm [Fe/H]}}{\sigma_{{\rm [Fe/H]}}} \right]^2 +
\left[ \frac{\Delta T_{\rm eff}}{\sigma_{T_{\rm eff}}} \right]^2 +
\left[ \frac{\Delta \rho_\star}{\sigma_{\rho_\star}} \right]^2 +
\left[ \frac{\Delta M_V}{\sigma_{M_V}} \right]^2.
\end{equation}
Here, the $\Delta$ quantities denote the deviations between the
observed and calculated values at each point. The asymmetric error bar
in $\rho_\star$ was taken into account by using different values of
$\sigma_{\rho_\star}$ depending on the sign of the deviation. The
weights were further multiplied by a factor taking into account the
number density of stars along each isochrone, assuming a Salpeter mass
function. We used the same code as Torres et al.~(2008) and refer the
reader to that paper for further details.

Through this analysis, we found $M_\star =
1.263_{-0.047}^{+0.035}$~$M_\odot$, $R_\star =
1.446_{-0.067}^{+0.099}$~$R_\odot$, and a stellar age of
$3.18_{-0.68}^{+0.52}$~Gyr. The corresponding planetary mass and
radius were obtained by merging the results for the stellar properties
with the parameters determined in our analysis of the RV and
photometric data. The results are $M_p =
3.212_{-0.082}^{+0.069}$~$M_{\rm Jup}$ and $R_p =
1.023_{-0.055}^{+0.070}$~$R_{\rm Jup}$. These values are also given in
Table~\ref{tbl:params}, along with the values for some other
interesting parameters that can be derived from the preceding results.

As a consistency check, we computed the implied stellar surface
gravity and its uncertainty based on our analysis, finding $\log g =
4.219_{-0.055}^{+0.033}$ where $g$ is in cm~s$^{-2}$. This agrees with
the spectroscopic determination of surface gravity, $\log g = 4.29 \pm
0.06$ (Fischer et al.~2007), based on an analysis of the widths of
pressure-sensitive lines in the optical spectrum. Since the error bars
on the two results are comparable, one might be tempted to use the
spectroscopic $\log g$ as a further constraint on the stellar
properties. We did not take this approach out of concern that the
spectroscopic determination is more complex and liable to
underestimation of the error (see, e.g., Winn et al.~2008).

\section{Discussion}

\subsection{Summary and comparison with previous results}

We have presented photometry of a transit of HD~17156b, and 10
additional measurements of the radial velocity of the host star. We
have analyzed the data along with the other observed stellar
properties to refine the estimates of the basic system parameters. For
many of the parameters, our results are more precise than those
reported previously, despite some aspects of the previous analyses
that may have caused the quoted error bars to be unrealistically
small. Specifically, none of the previous analyses allowed for any
uncertainty in the limb-darkening law, and neither Barbieri et
al.~(2007) nor Irwin et al.~(2008) attempted to quantify the effect of
time-correlated noise. Another improvement in our analysis was the
integration of the data analysis with stellar evolution models to make
full use of the information in the light curve and arrive at a
self-consistent solution.

Our result for the stellar mass, $M_\star =
1.263_{-0.047}^{+0.035}$~$M_\odot$, agrees with the previous
determination of $1.2\pm 0.1$~$M_\odot$ by Fischer et al.~(2007) and
improves on the precision. Our result for the stellar radius,
$1.446_{-0.067}^{+0.099}$~$R_\odot$, agrees with the Fischer et
al.~(2007) estimate of $1.470\pm 0.085$~$R_\odot$. The essential
difference between the two analyses is that we made use of the
photometric determination of $\rho_\star$ while Fischer et al.~(2007)
used the spectroscopic determination of $\log g$. As for the planet,
our result of $1.023_{-0.055}^{+0.070}$~$R_{\rm Jup}$ is very similar
to the value $1.01\pm 0.09$~$R_{\rm Jup}$ found by Irwin et
al.~(2008). This should be interpreted as an agreement between the
measured transit depths, and not necessarily the other light curve
parameters, because Irwin et al.~(2008) did not fit for the
$a/R_\star$ parameter. Instead they used an external constraint on
that parameter based on the work by Fischer et al.~(2007). Gillon et
al.~(2008) also found the same transit depth, but a smaller value of
$a/R_\star$ and therefore larger values of $R_\star$ and $R_p$ (by
about 1$\sigma$, where $\sigma$ is the error quoted by Gillon et
al.~2008). Our result for $R_p$ is approximately 3 times more precise.

The results for the limb-darkening coefficients, given in
Table~\ref{tbl:ld}, show that the darkening is greater in the
$(b+y)/2$ band than in the $z$ band, as expected. The values of the
coefficients themselves are poorly constrained and the errors are
highly correlated, making it difficult to compare to specific
theoretical predictions. For the $z$ band, interpolation of the tables
of Claret~(2004) gives a prediction of $u_1=0.17$ and $u_2=0.35$,
which is just outside the 1$\sigma$ range of our results. The data
favor a less limb-darkened star. The tables of Claret~(2000) give
$u_1=0.52$ and $u_2=0.28$ for the $b$ band, and $u_1=0.39$ and
$u_2=0.33$ for the $y$ band. Either set of coefficients is compatible
with the loose bounds provided by the APT data.

\subsection{Comparison with theoretical models}

A primary goal of precise transit observations is to compare the
observed planetary properties with theoretical models of the planet's
interior structure. For example, a persistent theme in this field is
that at least a few planets have radii that are ``too large'' by the
standards of theoretical models of solar-composition giant planets,
even after accounting for the intense stellar heating and selection
effects (see Burrows et al.\ 2007 for a recent discussion). Other
planets are so small that the models fit only when the composition is
altered to be much richer in heavy elements than the Sun, a possible
indication of the dense interior cores that are expected according to
the core-accretion theory of planet formation (see, e.g., Sato et
al.~2005).

Bodenheimer et al.~(2003) give predictions for the radii of giant
planets as a function of the age, mass, and time-averaged equilibrium
temperature of the planet, defined as
\begin{equation}
T_{\rm eq} = \left[ \frac{(1-A) L_\star}{16\pi\sigma a^2(1+e^2/2)^2} \right]^{1/4} =~~(783~{\rm K})~(1-A)^{1/4},
\end{equation}
where $A$ is the Bond albedo, $L_\star$ is the stellar luminosity,
$\sigma$ is the Stefan-Boltzmann constant, $a$ is the semimajor axis,
and $e$ is the eccentricity. In the latter equality we have evaluated
$T_{\rm eq}$ for HD~17156b using the results given in
Table~\ref{tbl:params}. As long as the albedo is not very close to
unity, Bodenheimer et al.~(2003) predict a planetary radius of
1.10~$R_{\rm Jup}$ for a solar composition at 4.5~Gyr. They also
considered enriching the solar composition by 40~$M_\oplus$ of
additional heavy elements (4\% of the total mass) and found that the
radius decreases by less than 1\%.

Fortney et al.~(2007) have presented theoretical models parameterized
by mass, age, and an effective orbital distance, defined as the
distance from the Sun where a hypothetical planet on a circular orbit
would receive the same time-averaged flux as the actual planet,
\begin{equation}
d_\oplus = a(1+e^2/2) \left( \frac{L_\star}{L_\odot} \right)^{1/2} =~~0.13~{\rm AU},
\end{equation}
where again we have evaluated the expression as appropriate for
HD~17156b. Fortney et al.~(2007) also predicted $1.10$~$R_{\rm Jup}$
for a solar composition at 4.5~Gyr, decreasing to $1.02$~$R_{\rm Jup}$
when $100~M_\oplus$ of solar-composition material (10\% of the total
mass) is replaced by a heavy-element ``core.'' This would provide a
good match to the observed radius of 1.023~$R_{\rm Jup}$.

We conclude from these comparisons, as did Irwin et al.~(2008), that
HD~17156b is smaller than a theoretical giant planet of
solar-composition, and that this may be an indication of heavy-element
enrichment. Heavy-element enrichment is expected according to the
core-accretion model of planet formation (Mizuno 1980, Pollack et
al.~1996). There is evidence for such enrichment in Jupiter and
Saturn, with Jupiter in particular consisting of 3-15\% heavy elements
(Guillot 2005). One should be wary of over-interpretation, given that
our measurement of $R_p$ differs from 1.1~$R_{\rm Jup}$ by only 1.1
times the measurement uncertainty. However, there are some known
factors that would increase the theoretical radius and thereby enlarge
the discrepancy at least slightly: the fiducial radius calculated by
Bodenheimer et al.~(2003) and Fortney et al.~(2007) refers to a higher
pressure (smaller radius) than the transit-measured radius (Burrows et
al.~2003); the models do not take into account tidal heating and
consequent inflation due to the nonzero eccentricity (Liu et
al.~2008); and the age of the system is estimated to be 3~Gyr, younger
than the 4.5~Gyr age for which the models were calculated.

\subsection{The spin-orbit angle, and the probability of secondary
eclipses}

One parameter that we have not improved on, but that deserves mention,
is the angle between the stellar spin axis and the orbital axis. One
might expect the axes to be well-aligned, given that the star and
planet formed from a common disk and given the good alignment observed
in the Solar system. Then again, by the same logic the orbit should be
circular, and it is not. A lower limit on the angle between the spin
and orbital axes can be derived by monitoring the apparent Doppler
shift throughout a transit, exploiting the Rossiter-McLaughlin (RM)
effect (see, e.g., Queloz et al.~2000, Winn et al.~2005). Using this
technique, Narita et al.~(2008) found that the angle between sky
projections of the two axes is $\lambda = 62\pm 25$~deg, a 2.5$\sigma$
misalignment. Cochran et al.~(2008) presented two different data sets
giving $\lambda = 4.5\pm 15.6$~deg and $\lambda = -32.4\pm 25.2$~deg,
and concluded that the data are consistent with good
alignment. Unfortunately, our new data alone do not allow for
significant progress on this issue, because the limiting errors are in
the precision of the RM data, which we have not improved in this
study. Further spectroscopic observations are warranted.

Another important angle is the orbital inclination with respect to the
sky plane. If the orbit is oriented close enough to edge-on, then the
planet will be periodically eclipsed by the parent star.  Observations
of such secondary eclipses would reveal the planetary albedo or
thermal emission from the planet, depending upon the observing
bandpass. In addition, infrared observations could help to understand
the radiative dynamics of the planetary atmosphere, as emphasized by
Barbieri et al.~(2007). To check whether secondary eclipses are likely
to occur, we used our MCMC results to compute the {\it a posteriori}\,
probability distribution for the impact parameter at superior
conjunction,
\begin{equation}
b_{\rm II} = \frac{a \cos i}{R_\star} \left( \frac{1-e^2}{1-e\sin\omega} \right).
\end{equation}
We find the probability for secondary eclipses ($b_{\rm II} <
1+R_p/R_\star)$ to be 18\%.  The probability for complete eclipses, or
occultations ($b_{\rm II} < 1+R_p/R_\star$), is 15\%. These odds are
better than those found by Irwin et al.~(2008), which were 6.9\% and
9.2\%. Gillon et al.~(2008) found the probability of occultations to
be 0.04\%. Presumably this significant difference is attributable to
our finding of a more edge-on orbit ($i=86.2_{-0.8}^{+2.1}$~deg, as
opposed to $85.5_{-1.2}^{+1.9}$~deg from Gillon et al.~(2008). Our
error bars for $i$ are no smaller than those reported previously
because the achievable error in $i$ worsens rapidly as $i$ approaches
90~deg (Carter et al.~2008). To help in planning observations, we have
used our results to predict the timing of the events, as well as the
quantity $(R_p/d_{\rm II})^2$ (where $d_{\rm II}$ is the star-planet
distance at superior conjunction) which sets the amplitude of the
reflected-light signal from the planet. The results for this latter
parameter are conditioned on the assumption that secondary eclipses do
indeed occur. The results are given in
Table~\ref{tbl:superior-conjunction}. With a bit of luck, HD~17156b
will be eclipsed and give observers another gift.

\acknowledgments We are grateful to the referee, Kaspar von~Braun, for
a timely and detailed critique. This research was partly supported by
Grant No.~2006234 from the United States-Israel Binational Science
Foundation (BSF). Partial support also came from NASA Origins grants
NNG06GH69G (to M.J.H.) and NNG04LG89G (to G.T.). G.W.H.\ acknowledges
support from NASA, NSF, and the State of Tennessee through its Centers
of Excellence program. J.A.J.\ is a National Science Foundation
Astronomy and Astrophysics Postdoctoral Fellow with support from the
NSF grant AST-0702821. KeplerCam was developed with partial support
from the Kepler Mission under NASA Cooperative Agreement NCC2-1390 and
the Keplercam observations described in this paper were partly
supported by grants from the Kepler Mission to SAO and PSI.

\begin{deluxetable}{lccc}
\tabletypesize{\scriptsize}
\tablecaption{Predicted Superior Conjunction Parameters for HD~17156b\label{tbl:superior-conjunction}}
\tablewidth{0pt}

\tablehead{
\colhead{Parameter} & \colhead{Value} & \colhead{68.3\% Conf.~Limits}
}

\startdata
Midpoint of superior conjunction~[HJD]                    & $2454464.627$  &  $-0.090$, $+0.100$ \\
Probability of occultation, $P(b_{\rm II} < 1+R_p/R_\star)$   & 18\%        &  \nodata \\
Probability of non-grazing occultation, $P(b_{\rm II} < 1-R_p/R_\star)$     & 15\%  &  \nodata \\
Occultation duration for an edge-on orbit~[hr]        & $12.9$           &  $-0.9$,$+1.4$ \\
Reflected-light figure of merit, $10^6~(R_p/d_{\rm II})^2$  & $4.68$    &  $-0.22$, $+0.31$
\enddata

\tablecomments{The midpoint of superior conjunction has {\it not}\,
  been corrected for the light-travel time across the system.  The
  result for $(R_p/d_{\rm II})^2$ is conditioned on the occurrence of
  occultations.}

\end{deluxetable}

\end{document}

%% file: table_rv.tex
  1 &  $  2453746.75853$  &    $-$3&88  &  $   1.58$  \\
  1 &  $  2453748.80062$  &    43&59  &  $   1.75$  \\
  1 &  $  2453749.79720$  &    54&23  &  $   1.70$  \\
  1 &  $  2453750.80399$  &    73&76  &  $   1.71$  \\
  1 &  $  2453775.77928$  &   140&82  &  $   1.67$  \\
  1 &  $  2453776.80892$  &   161&32  &  $   1.55$  \\
  1 &  $  2453779.82980$  &   143&35  &  $   1.66$  \\
  1 &  $  2453959.13100$  &     3&25  &  $   1.52$  \\
  1 &  $  2453962.06926$  &    55&17  &  $   1.41$  \\
  1 &  $  2453963.10508$  &    72&97  &  $   1.50$  \\
  1 &  $  2453964.13028$  &    97&69  &  $   1.69$  \\
  1 &  $  2453982.03249$  &    39&66  &  $   1.12$  \\
  1 &  $  2453983.08599$  &    52&59  &  $   1.57$  \\
  1 &  $  2453983.99510$  &    70&80  &  $   1.25$  \\
  1 &  $  2453985.00883$  &    93&70  &  $   1.57$  \\
  1 &  $  2454023.95452$  &    15&98  &  $   1.84$  \\
  1 &  $  2454047.96100$  &    73&25  &  $   1.66$  \\
  1 &  $  2454083.90654$  &   $-$61&08  &  $   1.42$  \\
  1 &  $  2454084.83198$  &   $-$29&07  &  $   1.58$  \\
  1 &  $  2454085.86874$  &   $-$11&78  &  $   1.73$  \\
  1 &  $  2454129.92683$  &    20&22  &  $   1.35$  \\
  1 &  $  2454130.73184$  &    41&18  &  $   1.35$  \\
  1 &  $  2454131.85644$  &    53&68  &  $   1.83$  \\
  1 &  $  2454138.76840$  &   170&09  &  $   1.33$  \\
  1 &  $  2454319.12775$  &    $-$8&71  &  $   1.25$  \\
  1 &  $  2454336.07987$  &  $-$146&93  &  $   1.27$  \\
  1 &  $  2454337.12124$  &  $-$109&19  &  $   1.40$  \\
  1 &  $  2454339.13050$  &   $-$37&48  &  $   1.17$  \\
  1 &  $  2454427.82655$  &    40&82  &  $   1.45$  \\
  1 &  $  2454428.86486$  &    59&19  &  $   1.61$  \\
  1 &  $  2454545.72272$  &  $-$354&79  &  $   2.09$  \\
  1 &  $  2454545.72680$  &  $-$354&61  &  $   2.10$  \\
  1 &  $  2454546.82753$  &  $-$244&68  &  $   1.41$  \\
  1 &  $  2454546.83309$  &  $-$243&35  &  $   1.37$  \\
  2 &  $  2454078.01509$  &  $-$116&60  &  $   5.14$  \\
  2 &  $  2454078.92847$  &  $-$261&56  &  $   5.18$  \\
  2 &  $  2454079.91716$  &  $-$164&10  &  $   5.23$  \\
  2 &  $  2454080.98437$  &   $-$89&57  &  $   5.13$  \\
  2 &  $  2454081.89749$  &   $-$44&08  &  $   5.15$  \\
  2 &  $  2454082.86412$  &     2&39  &  $   5.14$  \\
  2 &  $  2454083.88785$  &    37&62  &  $   5.16$  \\
  2 &  $  2454085.82897$  &    86&67  &  $   5.20$  \\
  2 &  $  2454086.88295$  &    99&01  &  $   5.20$  \\

%% file: ms.bbl
\begin{thebibliography}{}

\bibitem[Agol et al.(2005)]{2005MNRAS.359..567A} Agol, E., Steffen,
  J., Sari, R., \& Clarkson, W.\ 2005, \mnras, 359, 567

\bibitem[Allan(1966)]{allan66} Allan, D.~W.\ 1966, Proc.\ IEEE, 54,
  221

\bibitem[Barbieri et al.(2007)]{2007A&A...476L..13B} Barbieri, M., et
  al.\ 2007, \aap, 476, L13

\bibitem[Barnes(2007)]{2007PASP..119..986B} Barnes, J.~W.\ 2007,
  \pasp, 119, 986

\bibitem[Bodenheimer et al.(2003)]{2003ApJ...592..555B} Bodenheimer,
  P., Laughlin, G., \& Lin, D.~N.~C.\ 2003, \apj, 592, 555

\bibitem[Burke(2008)]{2008ApJ...679.1566B} Burke, C.~J.\ 2008, \apj,
  679, 1566

\bibitem[Burrows et al.(2003)]{2003ApJ...594..545B} Burrows, A.,
  Sudarsky, D., \& Hubbard, W.~B.\ 2003, \apj, 594, 545

\bibitem[Burrows et al.(2007)]{2007ApJ...661..502B} Burrows, A.,
  Hubeny, I., Budaj, J., \& Hubbard, W.~B.\ 2007, \apj, 661, 502
 
\bibitem[Butler et al.(1996)]{1996PASP..108..500B} Butler, R.~P.,
  Marcy, G.~W., Williams, E., McCarthy, C., Dosanjh, P., \& Vogt,
  S.~S.\ 1996, \pasp, 108, 500

\bibitem[Carter et al.(2008)]{2008arXiv0805.0238C} Carter, J.~A., Yee,
  J.~C., Eastman, J., Gaudi, B.~S., \& Winn, J.~N.\ 2008, ArXiv
  e-prints, 805, arXiv:0805.0238

\bibitem[Charbonneau et al.(2007)]{2007prpl.conf..701C} Charbonneau,
  D., Brown, T.~M., Burrows, A., \& Laughlin, G.\ 2007, Protostars and
  Planets V, 701

\bibitem[Claret(2000)]{2000A&A...363.1081C} Claret, A.\ 2000, \aap,
  363, 1081

\bibitem[Claret(2004)]{2004A&A...428.1001C} Claret, A.\ 2004, \aap,
  428, 1001

\bibitem[Cochran et al.(2008)]{2008ApJ...683L..59C} Cochran, W.~D.,
  Redfield, S., Endl, M., \& Cochran, A.~L.\ 2008, \apjl, 683, L59

\bibitem[Demarque et al.(2004)]{2004ApJS..155..667D} Demarque, P.,
  Woo, J.-H., Kim, Y.-C., \& Yi, S.~K.\ 2004, \apjs, 155, 667

\bibitem[Dravins et al.(1998)]{1998PASP..110..610D} Dravins, D.,
  Lindegren, L., Mezey, E., \& Young, A.~T.\ 1998, \pasp, 110, 610

\bibitem[Fischer et al.(2007)]{2007ApJ...669.1336F} Fischer, D.~A., et
  al.\ 2007, \apj, 669, 1336

\bibitem[Fortney et al.(2007)]{2007ApJ...659.1661F} Fortney, J.~J.,
  Marley, M.~S., \& Barnes, J.~W.\ 2007, \apj, 659, 1661

\bibitem[Gillon et al.(2008)]{2008A&A...485..871G} Gillon, M., Triaud,
  A.~H.~M.~J., Mayor, M., Queloz, D., Udry, S., \& North, P.\ 2008,
  \aap, 485, 871

\bibitem[Guillot(2005)]{2005AREPS..33..493G} Guillot, T.\ 2005, Annual
  Review of Earth and Planetary Sciences, 33, 493

\bibitem[H{\o}g et al.(2000)]{2000A&A...355L..27H} H{\o}g, E., et al.\
  2000, \aap, 355, L27

\bibitem[Holman \& Murray(2005)]{2005Sci...307.1288H} Holman, M.~J.,
  \& Murray, N.~W.\ 2005, Science, 307, 1288

\bibitem[Holman et al.(2006)]{2006ApJ...652.1715H} Holman, M.~J., et
  al.\ 2006, \apj, 652, 1715

\bibitem[Irwin et al.(2008)]{2008ApJ...681..636I} Irwin, J., et al.\
  2008, \apj, 681, 636

\bibitem[Kane \& von Braun(2008)]{2008arXiv0808.1890K} Kane, S.~R., \&
  von Braun, K.\ 2008, arXiv:0808.1890

\bibitem[Kibrick et al.(2006)]{2006SPIE.6274E..58K} Kibrick, R.~I.,
  Clarke, D.~A., Deich, W.~T.~S., \& Tucker, D.\ 2006, \procspie,
  6274, 62741U

\bibitem[Langton \& Laughlin(2008)]{2008ApJ...674.1106L} Langton, J.,
  \& Laughlin, G.\ 2008, \apj, 674, 1106

\bibitem[Liu et al.(2008)]{2008arXiv0805.1733L} Liu, X., Burrows, A.,
  \& Ibgui, L.\ 2008, ArXiv e-prints, 805, arXiv:0805.1733

\bibitem[Mandel \& Agol(2002)]{2002ApJ...580L.171M} Mandel, K., \&
  Agol, E.\ 2002, \apjl, 580, L171

\bibitem[Mizuno(1980)]{1980PThPh..64..544M} Mizuno, H.\ 1980, Progress of 
Theoretical Physics, 64, 544 

\bibitem[Narita et al.(2008)]{2008PASJ...60L...1N} Narita, N., Sato,
  B., Ohshima, O., \& Winn, J.~N.\ 2008, \pasj, 60, L1

\bibitem[P{\'a}l(2008)]{2008arXiv0805.2157P} P{\'a}l, A.\ 2008, ArXiv
  e-prints, 805, arXiv:0805.2157

\bibitem[Pollack et al.(1996)]{1996Icar..124...62P} Pollack, J.~B.,
  Hubickyj, O., Bodenheimer, P., Lissauer, J.~J., Podolak, M., \&
  Greenzweig, Y.\ 1996, Icarus, 124, 62

\bibitem[Press et al.(1992)]{nr} Press, W.~H., Teukolsky, S.~A.,
  Vetterling, T., \& Flannery, B.~P.~1992, Numerical Recipes in C
  (Cambridge: Cambridge Univ. Press)

\bibitem[Queloz et al.(2000)]{2000A&A...359L..13Q} Queloz, D.,
  Eggenberger, A., Mayor, M., Perrier, C., Beuzit, J.~L., Naef, D.,
  Sivan, J.~P., \& Udry, S.\ 2000, \aap, 359, L13

\bibitem[Reiger(1963)]{1963AJ.....68..395R} Reiger, S.~H.\ 1963, \aj,
  68, 395

\bibitem[Sato et al.(2005)]{2005ApJ...633..465S} Sato, B., et al.\
  2005, \apj, 633, 465

\bibitem[Seager \& Mall{\'e}n-Ornelas(2003)]{2003ApJ...585.1038S}
  Seager, S., \& Mall{\'e}n-Ornelas, G.\ 2003, \apj, 585, 1038

\bibitem[Seager(2008)]{2008SSRv..tmp...11S} Seager, S.\ 2008, Space
  Science Reviews, 11

\bibitem[Southworth(2008)]{2008MNRAS.386.1644S} Southworth, J.\ 2008, 
  \mnras, 386, 1644 

\bibitem[Sozzetti et al.(2007)]{2007ApJ...664.1190S} Sozzetti, A.,
  Torres, G., Charbonneau, D., Latham, D.~W., Holman, M.~J., Winn,
  J.~N., Laird, J.~B., \& O'Donovan, F.~T.\ 2007, \apj, 664, 1190

\bibitem[Szentgyorgyi et al.(2005)]{2005AAS...20711010S} Szentgyorgyi,
  A.~H., et al.\ 2005, Bulletin of the American Astronomical Society,
  37, 1339

\bibitem[Tegmark et al.(2004)]{2004PhRvD..69j3501T} Tegmark, M., et
  al.\ 2004, \prd, 69, 103501

\bibitem[Thompson et al.(2001)]{2001isra.book.....T} Thompson, A.~R.,
  Moran, J.~M., \& Swenson, G.~W., Jr.\ 2001, Interferometry and
  synthesis in radio astronomy, 2nd ed.\ (New York: Wiley)

\bibitem[Torres et al.(2008)]{2008ApJ...677.1324T} Torres, G., Winn,
  J.~N., \& Holman, M.~J.\ 2008, \apj, 677, 1324

\bibitem[Van Leeuwen(2007)]{2007ASSL..250.....V} van Leeuwen, F.\
  2007, in Hipparcos, the New Reduction of the Raw Data, Astrophysics
  and Space Science Library, Vol.\ 350 (Dordrecht: Springer)

\bibitem[Vogel(1890)]{1890AN....123..289V} Vogel, H.~C.\ 1890,
  Astronomische Nachrichten, 123, 289

\bibitem[Vogt et al.(1994)]{1994SPIE.2198..362V} Vogt, S.~S., et al.\ 1994, 
  \procspie, 2198, 362 

\bibitem[Winn et al.(2005)]{2005ApJ...631.1215W} Winn, J.~N., et al.\
  2005, \apj, 631, 1215

\bibitem[Winn et al.(2007)]{2007ApJ...657.1098W} Winn, J.~N., Holman,
  M.~J., \& Roussanova, A.\ 2007, \apj, 657, 1098

\bibitem[Winn(2008)]{w08-iau253} Winn, J.N.~2008, to appear in Proc.\
  IAU Symp.\ No.\ 253, ``Transiting Planets,'' eds.\ F.~Pont et
  al. (Cambridge Univ.\ Press)

\bibitem[Winn et al.(2008)]{2008arXiv0804.4475W} Winn, J.~N., et al.\
  2008, ApJ, 683, 1076

\bibitem[Wright(2005)]{2005PASP..117..657W} Wright, J.~T.\ 2005,
  \pasp, 117, 657

\bibitem[Yi et al.(2001)]{2001ApJS..136..417Y} Yi, S., Demarque, P.,
  Kim, Y.-C., Lee, Y.-W., Ree, C.~H., Lejeune, T., \& Barnes, S.\
  2001, \apjs, 136, 417

\bibitem[Young(1967)]{1967AJ.....72..747Y} Young, A.~T.\ 1967, \aj,
  72, 747


\end{thebibliography}
